**Inertial Motion in the Events Plane of Minkowski Space with**

**Non-zero Rest Mass (Axiomatic Description)**

Isaac Vagner

**Abstract**

The Lorentz group is a fundamental of an events space, in which physical processes for which gravitation is of no essential importance are considered. From a kinematics point of view, these processes are characterized by corresponding geometrical figures, formed by world lines of the particles that take part in the process. The geometric properties of the figures reflect peculiarities of the processes. This work deals with two-dimensional space in the case of inertial motion.

Inertial motion is considered in the plane of events characterized by the homogeneous Lorentz group L. On the basis of this group, a set of inertial movements and its decomposition into sets which are disconnected from one another with respect to the L-subgroups are considered. The geometric and corresponding physical characteristics of these motions are discussed: relativistic velocity, mass, relativistic momentum and mass/velocity ratio. It is shown that only one world line of inertial motion corresponds to each point on the plane in a space-like area, and the mass growth, dependent on the velocity, takes place only in the particle system.

The axiomatic description corresponds to principles of geometric construction, based only on group theory.

**Keywords**: Lorentz group, Minkowski space, inertial motion world lines, inertial rest mass, non–interacting particles.

## 1. Introduction

Inertial motion is described in the plane of events space on the basis of the law of inertia. The properties of the event plane are expressed by the homogeneous two-dimensional Lorentz group:

$$
\text{L:} \qquad \left.\begin{aligned} x' &= x \cdot \cosh\theta + \tau \cdot \sinh\theta \\ \tau' &= x \cdot \sinh\theta + \tau \cdot \cosh\theta \end{aligned}\right\} \tag{1}
$$

where the hyperbolic angle $\theta$ is the transformation parameter. The relation of $\theta$ to the relative velocity $\beta$ is given by $\beta = tanh\,\theta$. The coordinate $\boldsymbol{\tau}$ = ct, where t is the time, and c is the maximal velocity, the speed of light. In $(\boldsymbol{x}, \boldsymbol{\tau})$ coordinates, c = 1. The invariant of the transformations (1) is

$$-x^2 + \tau^2 = R^2 = \text{inv} \tag{2}$$

The orthogonal basis $\{e_1, e_2\}$ corresponding to it is

$$e_1^2 = -1, \quad e_2^2 = 1, \qquad e_1 e_2 = 0 \tag{3}$$

A vector in the plane of events is

$$\bar{r} = \mathrm{x}e_1 + \tau e_2 \tag{4}$$

The orthogonality condition for two vectors reads

$$\bar{r}_1 \bar{r}_2 = -\mathrm{x}_1\mathrm{x}_2 + \tau_1\tau_2 = 0 \tag{5}$$

The space of the group (1) is not homogeneous. This follows from the fact that the point (0,0) is a fixed point of the group L. The whole set of points in the events plane is divided into subsets of equivalent points with respect to L. The lines on which equivalent points are situated are called *orbits*. A set of orbits of the plane is described by relation (2) with parameter R. Group transformations move points only along the orbits of the group.

In Figure 1 one can see an orbit set of the events plane; the direction of the displacement of the points with $\theta > 0$ is shown by arrows.

Equation (2), by setting R = 0, defines four orbits $OC_1$, $OC_2$, $OC_3$ and $OC_4$ with slopes $K_X$ = $\pm$ 1. The (0,0) point is not included in these orbits. The orbit labeled 'n–n' in Figure 1 is a Minkowski circle with $R^2 > 0$, i.e. with a real radius R; the orbit in Figure 1 labeled 'm–m' is a circle with $R^2 < 0$; i.e., with imaginary radius.

## 2. Inertial motion in the events plane

The world line of inertial motion is a straight line. For an affine space, this straight line is characterized by the equation

$$a_1 x - a_2 \tau = a \tag{6}$$

where $a_1$, $a_2$, $a$ are coordinates of the straight line. Now let us analyze the peculiarities of the straight lines as we pass from the affine space to the space that is defined by the L group. For this purpose, using the transformation (1), we find the equations for coordinate transformations of the straight line (6).

$$\left.\begin{aligned} a_1' &= a_1 \cosh\theta + a_2 \sinh\theta \\ a_2' &= a_1 \sinh\theta + a_2 \cosh\theta \\ a' &= a \end{aligned}\right\} \tag{7}$$

The following invariants correspond to these transformations

$$\left.\begin{aligned} i_1 &= a = inv \\ i_2 &= -a_1^2 + a_2^2 = inv \end{aligned}\right\} \tag{8}$$

Due to their homogeneity, these invariants have no geometrical sense. They will acquire geometrical sense either in case of

$$i_1 = 0,\ i_2 = 0 \tag{9}$$

or by exclusion of homogeneity in the following values ratio

$$i_3 = \frac{i_1}{\sqrt{i_2}} = \frac{a}{\sqrt{-a_1^2+a_2^2}} \tag{10}$$

Let us define the sets of equivalent straight lines, corresponding to these invariants, and the geometrical meaning of the invariants.

Let us begin with the invariant $i_1 = a = 0$. Then (6) is written

$$\mathrm{x} = \frac{a_2}{a_1}\tau = K_x\tau \tag{11}$$

This is a bundle of straight lines with its center at (0,0). The only coordinate of this bundle is the straight line's slope $K_X$ $=\tanh\varphi$, where $\varphi$ is the hyperbolic angle of the slope. By means of the transformation (7) we can determine the formula of the $K_X$ transformation

$$K_x' = \frac{a_2'}{a_1'} = \frac{\sinh\theta + K_x\cosh\theta}{\cosh\theta + K_x\sinh\theta} = \tanh(\varphi+\theta) \tag{12}$$

Upon undergoing the transformation L, the straight lines of the bundle turn at an angle $\theta$.

According to the structure of the events plane (Fig.l), the straight lines of the bundle are situated in two disconnected sectors $C_1OC_2$, and $C_1OC_4$. In the $C_1OC_2$ sector, $|K_x| < 1$. We may use Eq.(12) to pass from the projective transformation to the linear transformation

$$\sqrt{1-K_x'^{\,2}} = \frac{\sqrt{1-K_x^2}}{\cosh\ \theta + K_x \sinh\ \theta} \tag{13}$$

In this case, according to (12) and (13),

$$\left.\begin{aligned} \frac{K_x'}{\sqrt{1-K_x'^{\,2}}} &= \frac{K_x}{\sqrt{1-K_x^2}}\cosh\theta + \frac{1}{\sqrt{1-K_x^2}}\sinh\theta \\ \frac{1}{\sqrt{1-K_x'^{\,2}}} &= \frac{K_x}{\sqrt{1-K_x^2}}\sinh\theta + \frac{1}{\sqrt{1-K_x^2}}\cosh\theta \end{aligned}\right\} \tag{14}$$

The transformations (14) are the transformations of the contravariant vector $\bar{u}$ in the events plane:

$$\bar{u} = \frac{K_x e_1 + e_2}{\sqrt{1-K_x^2}} = e_1 \sinh\varphi + e_2 \cosh\varphi = u_1 e_1 + u_2 e_2 \qquad (15)$$

The relative velocity of motion $K_x = \tanh\varphi$ is

$$\frac{u^1}{u^2} = K_x \qquad (16)$$

Its module is

$$|\bar{u}| = 1 \qquad (17)$$

Therefore $\bar{u}$ is the unit vector of a straight line of the bundle, situated in $C_1OC_2$ sector, where $|K_x| < 1$. Thus the vector $\bar{u}$ is the character of straight lines of a bundle with the center in (0,0) point. It is called the *relativistic velocity*. The straight lines bundle, situated in the $C_1OC_4$ sector, is characterized by the identical unit vector $\bar{u}_\perp$ (Fig. 1).

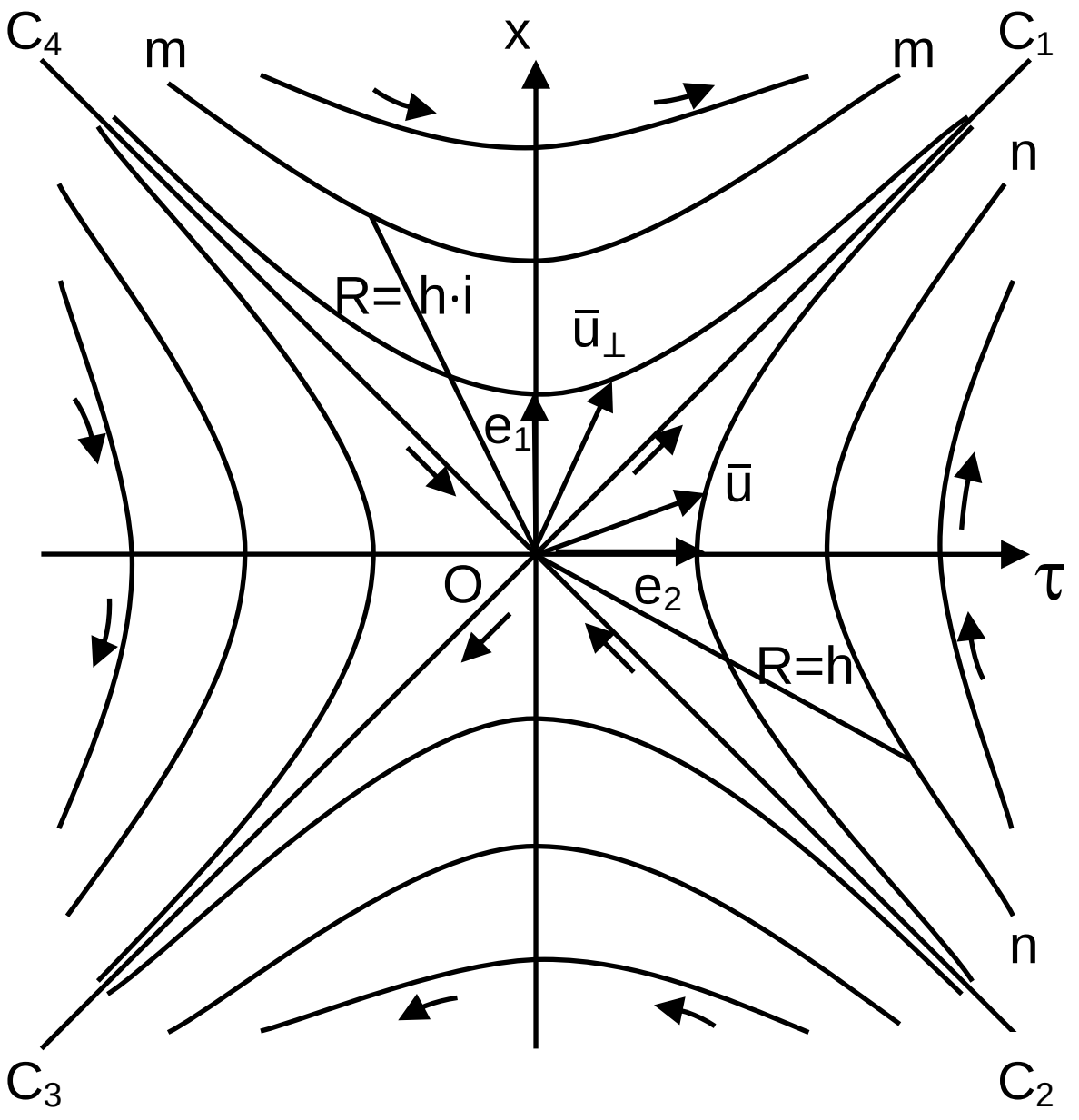

**Fig. 1**

**The structure of the events plane**.

It can be expressed by the $K_x$ slope of the vector $\bar{u}$:

$$\bar{u}_{\perp} = \frac{e_1 + K_x e_2}{\sqrt{1-K_x^2}} \tag{18}$$

$|\bar{u}_{\perp}| = -1$. This vector is orthogonal to vector $\bar{u}$. Therefore these two vectors together form an orthogonal basis (Fig. l).

The invariant $i_2 = 0$ is not analyzed here because it refers to the inertial motion of particles with $m_0 = 0$.

### 3. Inertial motion world lines not containing the point (0,0)

Let us consider the geometrical sense of the invariant $i_3$ (Eq.10) when $a \neq 0$, $a_1 \neq 0$, and $|K_x| < 1$. Then equation (6) becomes

$$\boldsymbol{x} = \boldsymbol{K_x \tau} + \boldsymbol{b} \tag{19}$$

where $K_x = \frac{a_2}{a_1}$, and $b = \frac{a}{a_1}$. According to (10)

$$i_3 = \frac{a}{\sqrt{-a_1^2 + a_2^2}} = \frac{\pm ib}{\sqrt{1-K_x^2}} = \pm ih = \text{inv} \tag{20}$$

By means of the transformations (7) we shall define the transformations of the $K_x$ and $b$ coordinates.

$$K_x' = \frac{a_2'}{a_1'} = \frac{\sinh\theta + K_x \cosh\theta}{\cosh\theta + K_x \sinh\theta} \tag{21}$$

$$b' = \frac{a'}{a_1'} = \frac{b}{\cosh\theta + K_x \sinh\theta} \tag{22}$$

The transformation (21) coincides with (12); therefore the set of straight lines, defined by the invariant $i_3$, has the same guiding vector $\bar{u}$ in the case of inertial motion $\bar{u} = \overline{OE}$ (Fig. 2), where $n_1 - n_1$ is a unit radius circle. AB is the world line of inertial motion.

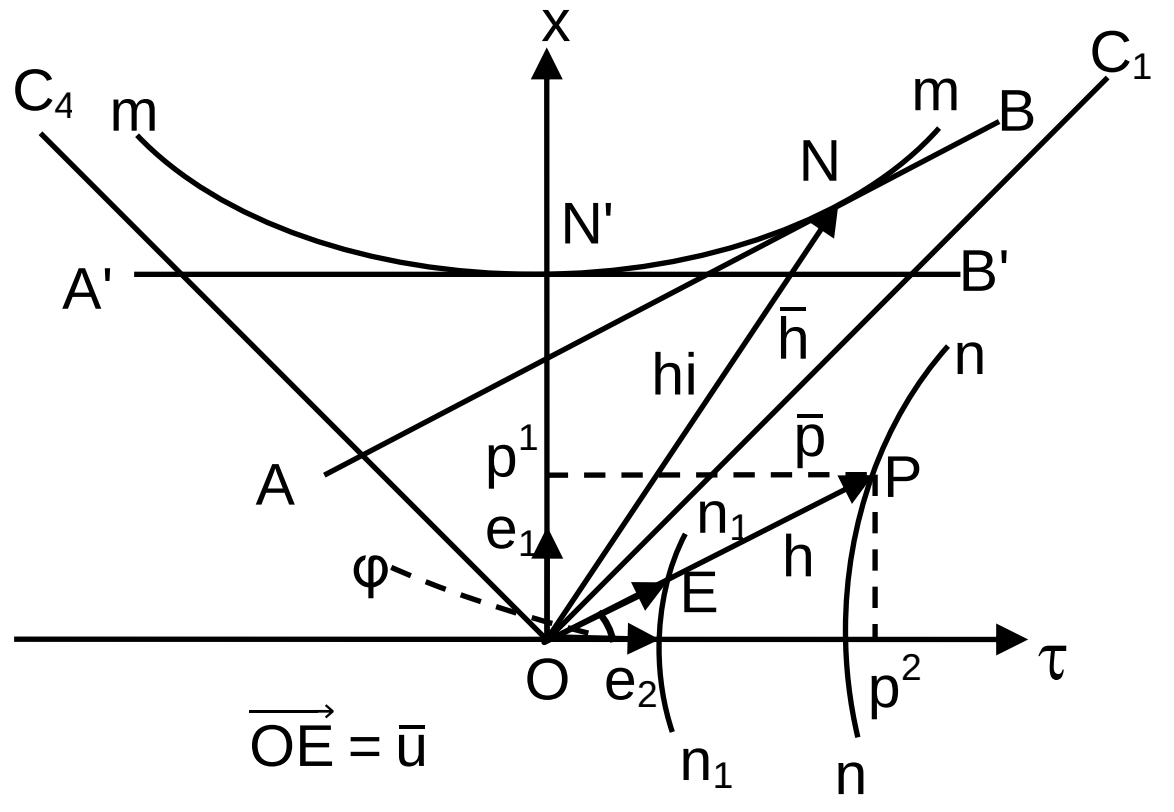

**Fig. 2**

**The subgroup of the world lines of inertial motion is given by the straight line AB. $\overline{\mathbf{OP}}$ is the representative vector of the straight line AB.**

To determine the geometric essence of the invariant $h$, we shall rotate the aforementioned world line AB $\rightarrow$ A'B' by means of transformations (21) in such a way that this line becomes parallel to the O$\tau$ axis (Fig. 2). Then $K_x' = 0$, and AB is determined by the equation $x' = b'$. According to (21), the rotation angle $\theta$ is determined by the relation $\tanh\theta = -K_X$ . Then, according to (22) and (20)

$$b' = \frac{b}{\sqrt{1-K_x^2}} = h = \text{inv} \tag{23}$$

As all straight lines in the $C_1OC_4$ sector have an imaginary length, the vector $\bar{h} = \overline{ON}'$ has the length $\boldsymbol{hi}$. It is a radial vector from the center O of the m–m circle (Fig. 2) to the point N', and it is orthogonal to the straight line A'B', which is tangent to the circle. During the transformations, the point N$'$ moves along its orbit m–m. The straight line A'B' moves together with this point and continues to be tangent to the circle; vector $\bar{h}$ moves also. As a result, we come to the conclusion that the invariant $\boldsymbol{hi}$ expresses the value of the radius of the circle which is the envelope of the family of equivalent straight lines, defined by some member AB of the family. During all transformations, the vector $\bar{h}$ remains orthogonal to the straight lines belonging to this family of lines. Therefore it is a polar radius to these straight lines.

While determining the characteristics of inertial motion world lines, it is advisable to pass from an invariant, expressed by an imaginary number $\boldsymbol{hi}$, to another invariant, expressed by the real number $h$, which is the length of the vector $\bar{P}$, symmetrical to $\bar{h}$ relative to $OC_1$:

$$\bar{P} = \overline{OP} = \mathrm{h}\bar{u} = p^1 e_1 + p^2 e_2 \tag{24}$$

According to Figure 2

$$p^1 = hu^1 = h\frac{K_x'}{\sqrt{1-K_x^2}} = h{\cdot}\sinh\varphi, \qquad p^2 = hu^2 = h\frac{1}{\sqrt{1-K_x^2}} = h\cdot\cosh\varphi \tag{25}$$

Vector $\bar{P}$ completely determines the family of equivalent straight lines. It is collinear with these straight lines, and its length equals $h$. Therefore we shall call this vector the *representative vector* of the family of straight lines. When $m_0 \neq 0$, all the representative vectors of the inertial motion are situated in the time-identical $C_1CO_2$ sector.

The physical meaning of the invariant $h$ will be analyzed later. I have only mentioned here that this is an inertial rest mass.

## 4. The system of non-interacting particles

Let us consider a system of two non–interacting particles, represented by world lines $A_1B_1$ and $A_2B_2$ (Fig. 3).

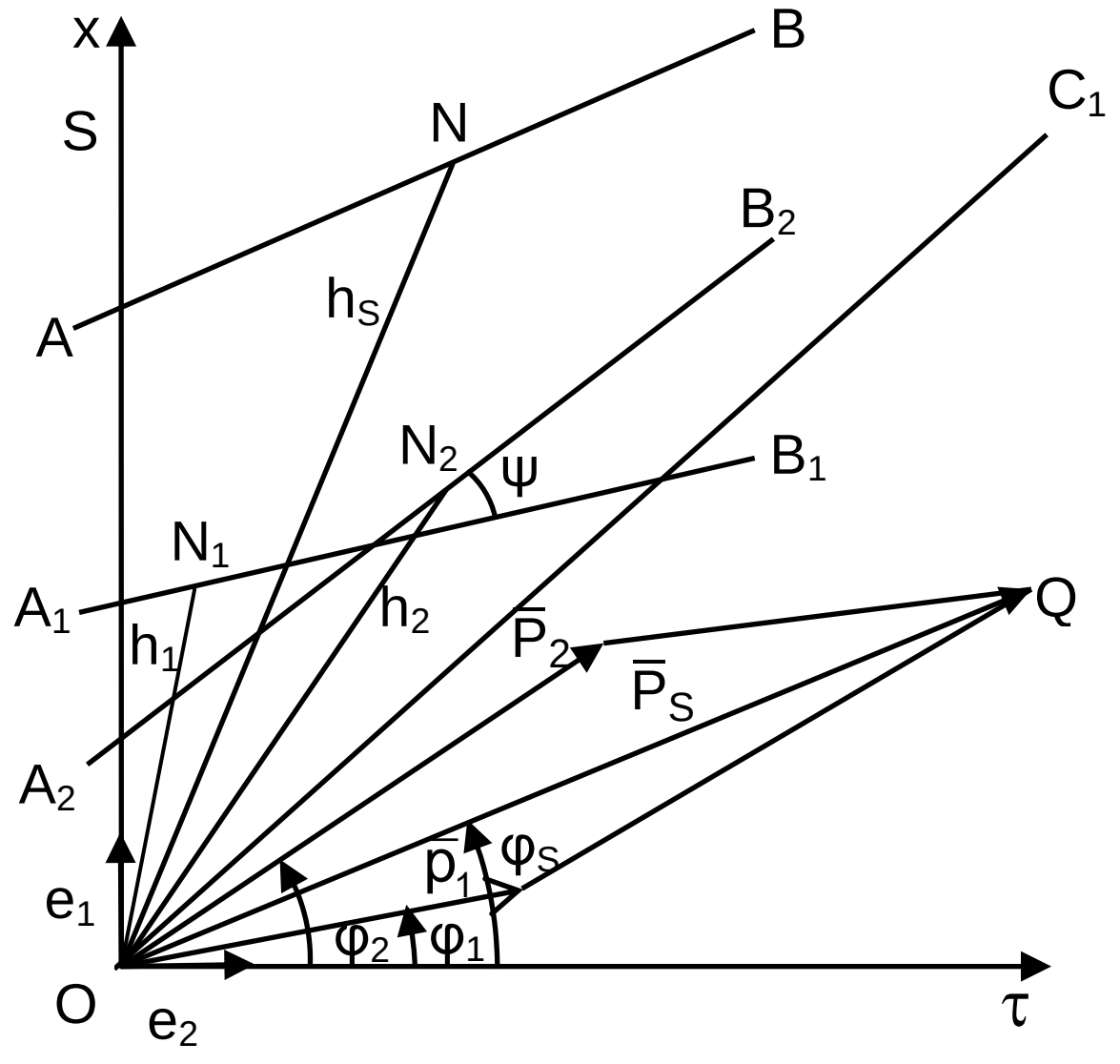

**Fig. 3**

**The system of two non-interacting particles. $\bar{P}_S$ is the representative vector of the system. The lines ON and OQ are symmetrical relative to $OC_1$.**

They are characterized by the representative vectors

$$\left.\begin{aligned} \bar{P}_1 &= h_1\bar{u}_1 = h_1\frac{K_{x1}e_1+e_1}{\sqrt{1-K_{x1}^2}} = h_1(e_1\sinh\varphi_1 + e_2\cosh\varphi_1) \\ \bar{P}_2 &= h_2\bar{u}_2 = h_2\frac{K_{x2}e_2+e_2}{\sqrt{1-K_{x2}^2}} = h_2(e_2\sinh\varphi_2 + e_2\cosh\varphi_2) \end{aligned}\right\} \qquad (26)$$

These straight lines intersect one another; this is a necessary condition of interaction between the particles according to the close-range conception. At the crossing point there is no break. This proves the absence of an interaction. In addition to the invariants $h_1$ and $h_2$, characterizing each particle separately, this system is described by a common invariant, the hyperbolic angle ψ between the straight lines:

$$\psi = \varphi_2 - \varphi_1 = artanh \frac{K_{x2}-K_{x1}}{1-K_{x1}K_{x2}} = \text{inv} \quad (27)$$

where $K_{x1}$ = tanh $\varphi_1$, $K_{x2}$ = tanh $\varphi_2$, are slopes for world lines $A_1B_1$ and $A_2B_2$, resp.

As there is no external action on the system, this system as a whole performs an inertial motion. We shall characterize it by the representative vector $\bar{P}_S = h_S \bar{u}_S$ . According to (14), $\bar{P}_1$ and $\bar{P}_2$ are contravariant vectors; hence they are summed according to the parallelogram law $\bar{P}_S = \bar{P}_1 + \bar{P}_2$ . According to (26), the polar radius and the slope of the representative vector of the system are determined by the following relations:

$$h_S = \sqrt{h_1^2 + h_2^2 + 2h_1h_2 \cosh\psi} \quad (28)$$

$$K_S = \frac{h_1 \sinh\ \varphi_1 + h_2 \sinh\ \varphi_2}{h_1 \cosh\ \varphi_1 + h_2 \cosh\ \varphi_2} = \tanh\varphi_S \quad (29)$$

AB is the world line of the system of particles.

### 5. Equivalent states of the system of non-interacting particles

The representative vector $\bar{P}_S$ of the system of two non-interacting particles is simply defined by the vectors $\bar{P}_1$ and $\bar{P}_2$ (Fig 4). There is a corresponding subset of the equivalent representative vectors for each particle . They are radius-vectors, whose ends are situated on the circle $n_1 - n_1$ and $n_2 - n_2$ (Fig. 4); these correspond to the first and the second particles resp.

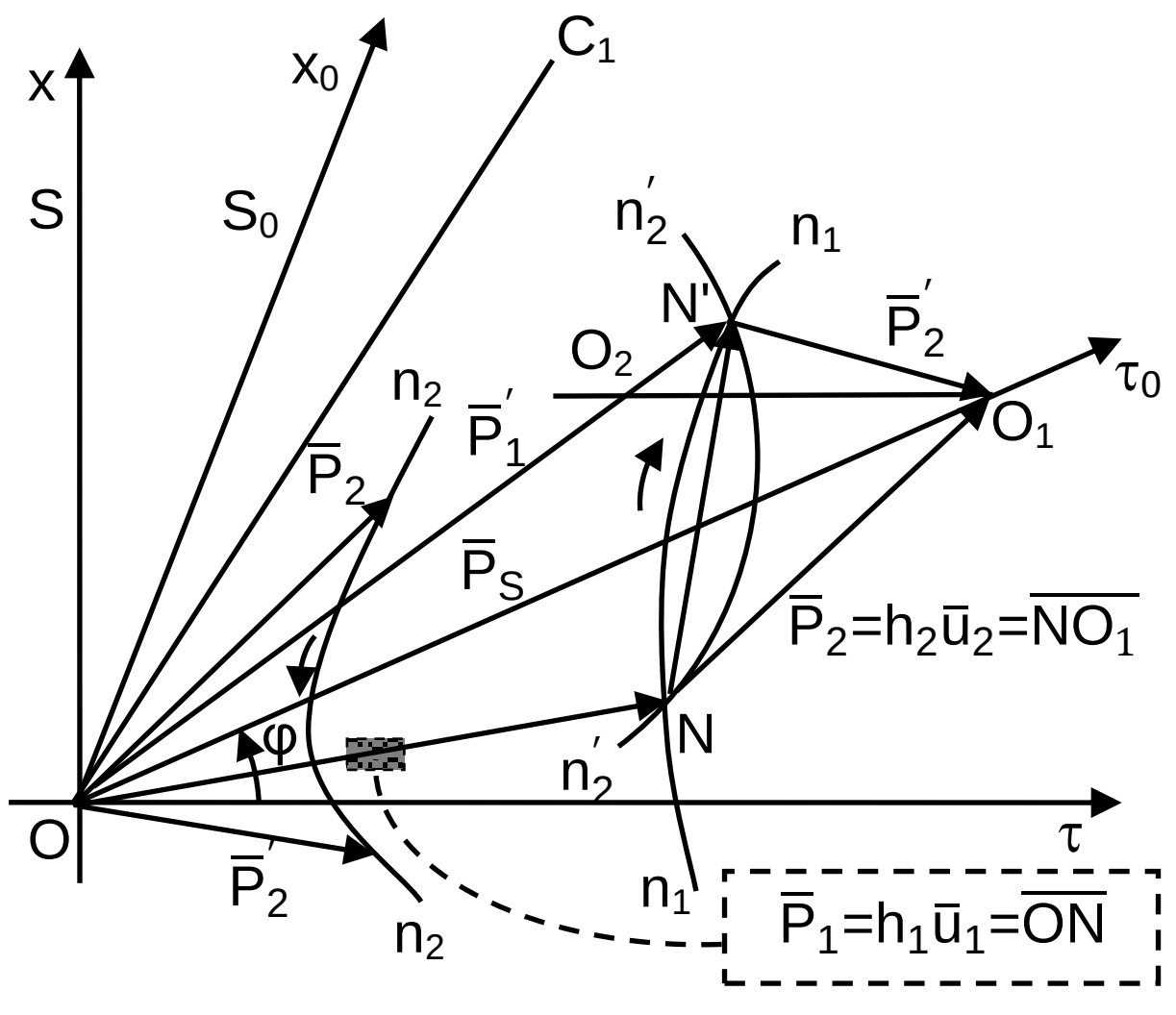

**Fig. 4**

**Equivalent states of the system of non-interacting particles. The same representative vector $\bar{P}_S$ corresponds to both systems.**

Within these subsets there are two pairs of vectors, $\bar{P}_1, \bar{P}_2,$ and $\bar{P}_1', \bar{P}_2'$ , whose sums define the same vector $\bar{P}_S = \bar{P}_1 + \bar{P}_2 = \bar{P}_1' + \bar{P}_2' = \overline{OO}_1$. The Minkowski circle whose center is at the point $O_1$, with radius $h_2$ ($\bar{P}_2 = h_2\bar{u}_2 = \overline{NO}_1$; the circle $n_2' - n_2'$ (Fig. 4)) meets the circle $n_1 - n_1$ at the two points N and N'. The sum of the vectors $\bar{P}_1$ and $\bar{P}_2$ corresponds to the point N. The point N' defines the second pair of the vectors $\bar{P}' = \overline{ON}'$, $\bar{P}_2' = \overline{N'O}_1$ . They are the respective mirror images of the vectors $\bar{P}_1$ and $\bar{P}_2$ with respect to the straight line $OO_1$, on which the vector $\bar{P}_S$ is situated. This symmetry is described by the Lorentz transformation:

$$\left.\begin{aligned} x' &= x \cdot \cosh 2\varphi + \tau \cdot \sinh 2\varphi \\ \tau' &= x \cdot \sinh 2\varphi + \tau \cdot \cosh 2\varphi \end{aligned}\right\} \qquad (30)$$

where $(x,\tau)$ , $(x',\tau')$ are the coordinates of the points N and N', resp., and $\varphi$ is the slope angle of the axis of symmetry $OO_1$. The transformations (30) do not change the lengths of the intervals. Therefore, the values of the polar radii remain unchanged.

$$h_1' = h_1, \quad h_2' = h_2 \tag{31}$$

Only relativistic velocities of the particles inside the system are changed. The total motion of the whole system remains unchanged. For this reason it is advisable to characterize both states of the system as equivalent states of the same system.

In the system of reference, in which the system of particles is at rest, the representative vector of the system coincides with the $O\tau$ axis (Fig.4). This is the system $S_0(x_0,\tau_0)$. It is named the *proper system of coordinates* for the system of particles.

## 6. Simple interaction

*Interaction* can be determined as a process leading to the change of the representative vector $\bar{P}$ of the moving object. This can be due either to the change of the relativistic velocity $\bar{u}$, leading to damage to the rectilinearity of the world line, or to a change of the value *h*. The latter is expressed by a parallel transfer of the world line that is not described by the homogeneous Lorentz group.

Let us consider the interaction of two particles. Before the interaction, these particles form a system of non-interacting particles with representative vectors $\bar{P}_1$ and $\bar{P}_2$ . Accordingly, the representative vector of the system is $\bar{P}_S = \bar{P}_1 + \bar{P}_2$. After the interaction, a new system of non-interacting particles is formed, with representative vectors $\bar{P}_1, \bar{P}_2$ , $\bar{P}_S = \bar{P}_1' + \bar{P}_2'$ . As a result, a transformation of the particles' vectors $\bar{P}_1 \to \bar{P}_1'$ , $\bar{P}_2 \to \bar{P}_2'$ takes place. The extent of the interaction should be defined by the extent of the deviation of world lines of particles from rectilinearity.

The interaction of particles takes place inside the system. The whole system as a unit is not subjected to external influences. Therefore this system performs an inertial movement with the same vector consistently through time:

$$\bar{P}_S = \bar{P}_1 + \bar{P}_2 = \bar{P}_1' + \bar{P}_2' \qquad (32)$$

We shall call this kind of interaction a *simple* interaction. Then a process of interaction from the space-time point of view is a discrete or a continuous succession of simple interactions unfolding in time.

To estimate the extent of the interaction of particles it is necessary to define the deviation of world lines of conforming particles from rectilinearity. We shall write relation (32) in the form $-(\bar{P}_1' - \bar{P}_1) = (\bar{P}_2' - \bar{P}_2)$ or

$$-(h_1'\bar{u}_1' - h_1\bar{u}_1) = (h_2'\bar{u}_2' - h_2\bar{u}_2) \qquad (33)$$

The left side of the equation (33) expresses the changes in the motion of the first particle, while the right side expresses the changes in the motion of the second particle.

## 7. The physical meaning of the invariant h. Mass relativistic impulse

To define the physical meaning of the invariant $h$, let us consider the interaction of two particles. Before the interaction, the particles are represented by the vectors $\bar{P}_1 = h_1\bar{u}_1$, and $\bar{P}_2 = h_2\bar{u}_2$ ; after the interaction they are represented by vectors $\bar{P}_1' = h_1'\bar{u}_1', and\ \bar{P}_2' = h_2'\bar{u}_2'$ .

The polar radii are invariants of group L. Therefore, in order to define physical meaning, we shall choose the interaction between particles during which the value $h$, does not change:

$$h_1 = h_1', \quad h_2 = h_2' \qquad (34)$$

Then from (33) follows

$$h_1 \Delta \bar{u}_1 = -h_2 \Delta \bar{u}_2 \qquad (35)$$

where $\Delta \bar{u}_1 = \bar{u}_1^{'} - \bar{u}_1$, $\Delta \bar{u}_2 = \bar{u}_2^{'} - \bar{u}_2$ are the changes in the relativistic velocities as a result of the interactions. At this time the common representative vector of the system is still unchanged.

Since condition (34) coincides with condition (31), and the vector $\bar{P}_S$ remains unchanged, particle systems before and after the interaction are equivalent with respect to the Lorentz group. Using the L-transformations, we shall rotate particles systems until $\bar{P}_S$ coincides with the Oτ axis (Fig. 5).

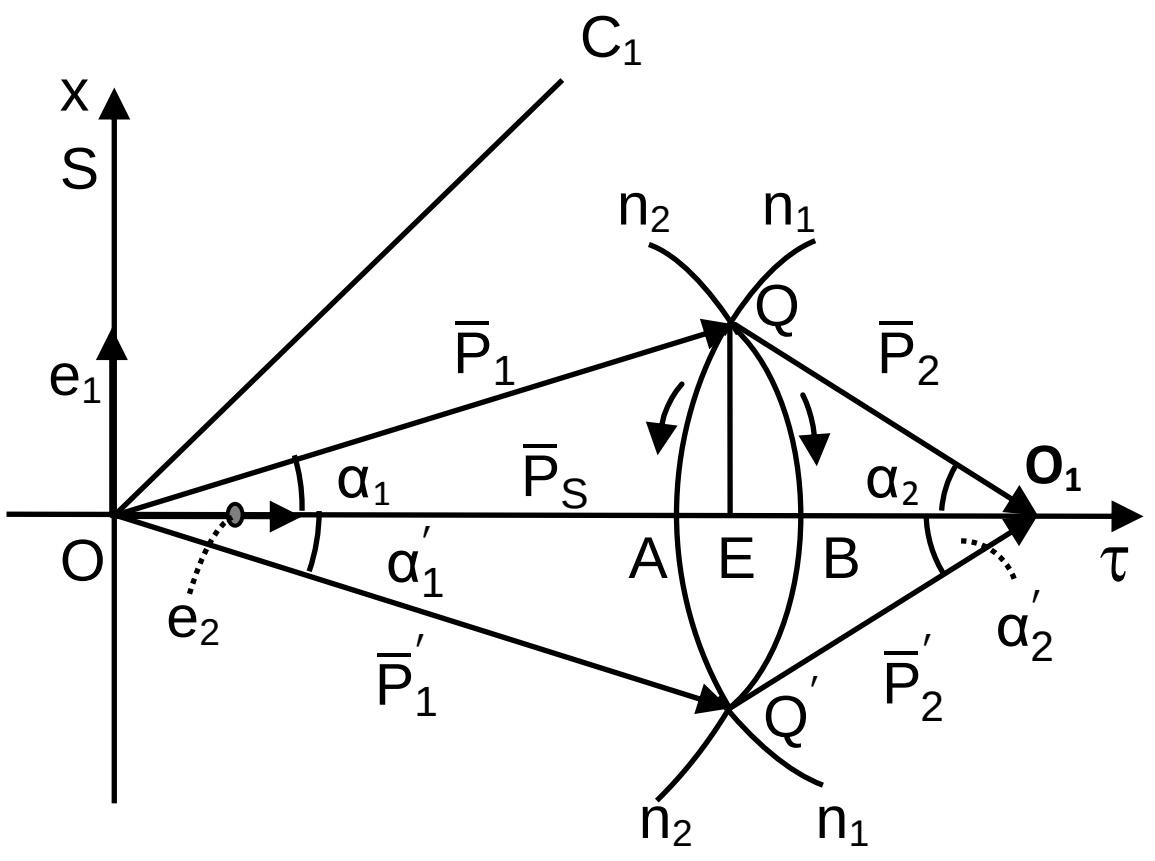

**Fig.5**

**The systems of non-interacting particles. Before the interaction, $\bar{P}_1 + \bar{P}_2 = \bar{P}_S$ ; after the interaction, $\bar{P}_1^{'} + \bar{P}_2^{'} = \bar{P}_S$. $m_{01}$ = OA and $m_{02} = O_1B$ are the rest masses of the first and second particles, resp. $m_1$ =OE and $m_2 = EO_1$ are the relativistic masses of the particles in the proper system S.**

**($\bar{P}_S$ coincides with Oτ.)**

Here the vectors $\bar{P}_1$ and $\bar{P}_2$ have the slopes $K_{\alpha_1} = \tanh \alpha_1$ and $K_{\alpha_2} = \tanh \alpha_2$ , while the vectors $\bar{P}_1^{'}$, $\bar{P}_2^{'}$ have the respective slopes

$$K_{\alpha_1}^{'} = K_{\alpha_1} , \qquad K_{\alpha_2}^{'} = K_{\alpha_2} \tag{36}$$

According to the foregoing, the vectors $\bar{P}_1^{'}$ and $\bar{P}_2^{'}$ are symmetrical with the vectors $\bar{P}_1$ and $\bar{P}_2$, relative to the Oτ axis, which coincides with the vector $\bar{P}_S$. We shall write the relativistic velocities of the particles before and after the interaction in the form derived from (15) combined with (36):

$$\left.\begin{aligned} \bar{u}_1 &= \frac{K_{\alpha_1} e_1 + e_2}{\sqrt{1-K_{\alpha_1}^2}} \quad \bar{u}_2 = \frac{-K_{\alpha_2} e_1 + e_2}{\sqrt{1-K_{\alpha_2}^2}} \\ \bar{u}_1^{'} &= \frac{-K_{\alpha_1} e_1 + e_2}{\sqrt{1-K_{\alpha_1}^2}} \quad \bar{u}_2^{'} = \frac{K_{\alpha_2} e_1 + e_2}{\sqrt{1-K_{\alpha_2}^2}} \end{aligned}\right\} \tag{37}$$

Accordingly, the changes in the relativistic velocities as a result of the interaction are

$$\Delta\bar{u}_1 = \bar{u}_1^{'} - \bar{u}_1 = \frac{\Delta K_{\alpha 1}}{\sqrt{1-K_{\alpha 1}^2}}\, e_1 \ , \quad \Delta\bar{u}_2 = \bar{u}_2^{'} - \bar{u}_2 = \frac{\Delta K_{\alpha 2}}{\sqrt{1-K_{\alpha 2}^2}}\, e_2 \tag{38}$$

where $\Delta K_{\alpha_1} = 2K_{\alpha_1}$ and $\Delta K_{\alpha_2} = 2K_{\alpha_2}$ are the changes of the relative velocities of the particles. As a result we can write the relation (35) as

$$\frac{h_1}{\sqrt{1-K_{\alpha 1}^2}}\, \Delta K_{\alpha_1} = \frac{h_2}{\sqrt{1-K_{\alpha 2}^2}}\, \Delta K_{\alpha_2} \tag{39}$$

or

$$m_1\, \Delta K_{\alpha_1} = m_2\, \Delta K_{\alpha_2} \tag{40}$$

where

$$m_i = \frac{h_i}{\sqrt{1-K_{\alpha_i}^2}}, \qquad \mathrm{i} = 1,2 \tag{41}$$

Since the angles $\alpha_1$ and $\alpha_2$ of the slopes are the angles between the vectors $\bar{P}_1, \bar{P}_S$ and $\bar{P}_2, \bar{P}_S$, the values of $m_i$ are invariants of the L-group. Thus the values of $m_i$ are the inertial motion characteristics. It follows from relation (40) that the higher the value of $m_i$, the less the change of the particle velocity $\Delta K_{\alpha_1}$ with the interaction under consideration. Thus the value of $m_i$ characterizes the particle's opportunity to resist the changing of the velocity of motion. This feature is called the *inertia* of the particle; the mass $m_i$ is the measure of inertia. It is called the *inertial mass* of the particle. In accordance with (41), the mass of the particle depends on its velocity $K_\alpha$ in the proper system and on the value of the polar radius. If $K_\alpha = 0$ then

$$m = h = m_0 \tag{42}$$

where $m_0$ is the rest mass of the particle. This forces us to conclude that the physical meaning of the invariant $h$ is essentially the rest mass of the moving object. According to (41)

$$m = \frac{m_0}{\sqrt{1-K_\alpha^2}} \tag{43}$$

The relation (43) expresses the dependence of the mass on the velocity of the particle while it is part of a whole system of particles that form the united moving object we are studying. We call $m$ the *relativistic mass*. According to (35), the interaction of particles is described by the equation

$$m_{01}\Delta\bar{u}_1 = -m_{02}\Delta\bar{u}_2 \tag{44}$$

while the motion of particles is characterized by the relativistic velocities $\Delta\bar{u}_1$ and $\Delta\bar{u}_2$.

Now we can define the physical essence of the representative vector $\bar{P}$.

$$\bar{P} = m_0\bar{u} = \frac{m_0 K_x}{\sqrt{1-K_x^2}}\, e_1 + \frac{m_0}{\sqrt{1-K_x^2}}\, e_2 \tag{45}$$

This vector is called the *relativistic momentum* of the particle. It expresses the velocity ($\bar{u}$) of the particle motion and its ability to resist changes in its motion ($m_0$).

The vector $\bar{P}$ is contravariant, so that its components, similarly to (14), are converted according to the formulas

$$\left.\begin{aligned} \frac{m_0 K_x'}{\sqrt{1-K_x'^{\,2}}} &= \frac{m_0 K_x}{\sqrt{1-K_x^2}}\cosh\theta + \frac{m_0}{\sqrt{1-K_x^2}}\sinh\theta \\ \frac{m_0}{\sqrt{1-K_x'^{\,2}}} &= \frac{m_0 K_x}{\sqrt{1-K_x^2}}\sinh\theta + \frac{m_0}{\sqrt{1-K_x^2}}\cosh\theta \end{aligned}\right\} \qquad (46)$$

where

$$\sinh\theta = \frac{\beta}{\sqrt{1-\beta^2}}\ , \qquad \cosh\theta = \frac{1}{\sqrt{1-\beta^2}} \qquad (47)$$

## 8. The relativistic mass of the particle system

Let us consider a system of two particles with relativistic momenta $\bar{P}_1' = m_{01}\bar{u}_1' = \overline{OQ}'$ , $\bar{P}_2' = m_{02}\bar{u}_2' = \overline{Q'O}'$ . $\bar{P}_S' = m_S\bar{u}_S' = \overline{OO}'$ is the relativistic momentum of the system (Fig. 6).

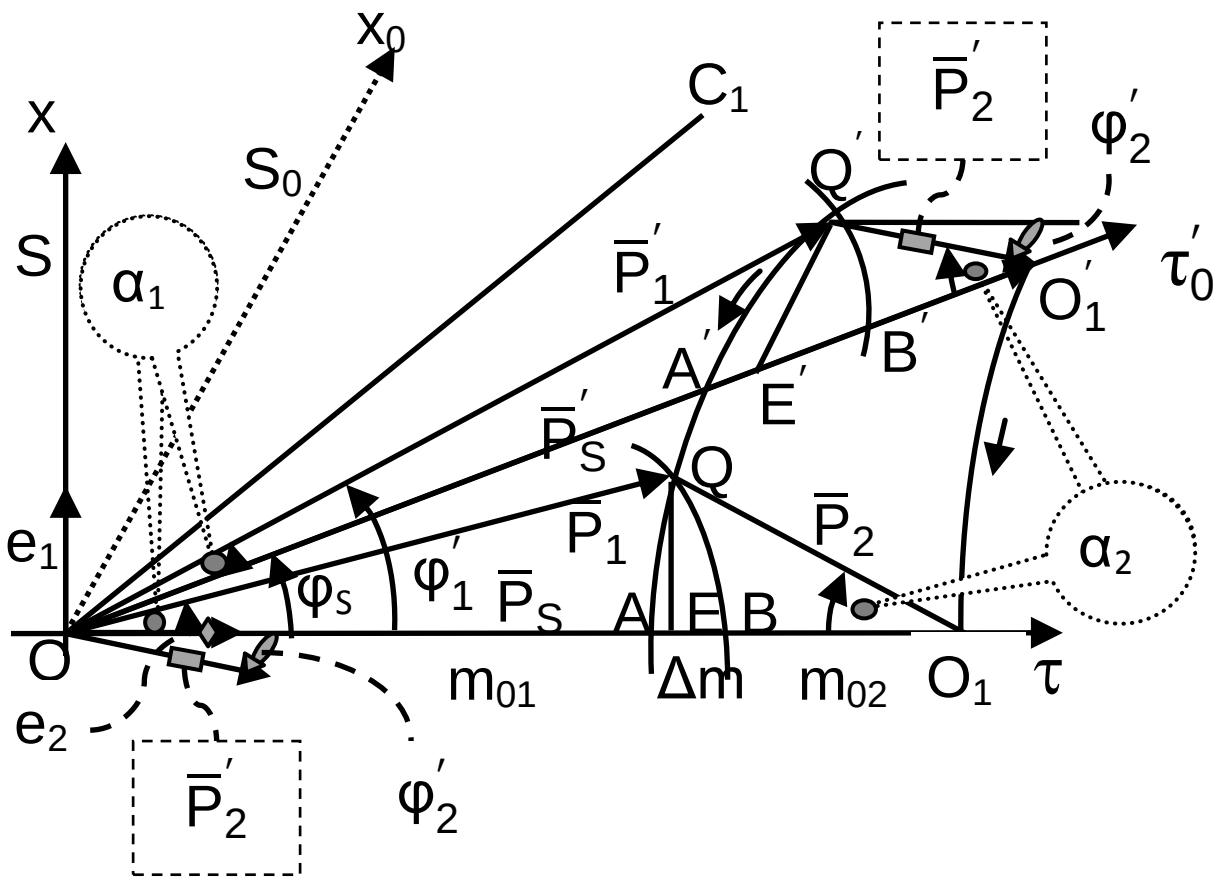

**Fig. 6**

**The systems of non-interacting particles in the proper systems of reference S and $S_0'$**

According to (28) and (29), the mass of the system $M_S$ and the slope $K_S^{'}$ of the vector $\bar{P}_S^{'}$ are defined as

$$M_S = \sqrt{m_{01}^2 + m_{02}^2 + 2m_{01}m_{02}\cosh\psi} = OO_1^{'} \quad (48)$$

$$K_S^{'} = \frac{m_{01}\sinh\varphi_1^{'} + m_{02}\sinh\varphi_2^{'}}{m_{01}\cosh\varphi_1^{'} + m_{02}\cosh\varphi_2^{'}} = \tanh\varphi_S^{'} \quad (49)$$

where $\varphi_1^{'}$, $\varphi_2^{'}$, $\varphi_S^{'}$ are the angles of the slopes of the vectors $\bar{P}_1, \bar{P}_2, \bar{P}_S$. , resp. $\psi = \varphi_2 - \varphi_1$. To analyze the mass of the system we shall rotate it in such a way that the vector $\bar{P}_S^{'}$ will coincide with the O$\tau$ axis; i.e. we shall study it in its own system of reference: $\bar{P}_S^{'} \to \bar{P}_S = \overline{OO}_1$.

Then $\bar{P}_1^{'} \to \bar{P}_1$ , $\bar{P}_2^{'} \to \bar{P}_2$ , $\varphi_1^{'} \to \varphi_1$ , $\varphi_2^{'} \to \varphi_2, \varphi_S^{'} = 0$ , $S_0^{'} \to S$. The condition of this rotation is $K_S^{'} = 0$ . According to (49)

$$m_{01}\sinh\alpha_1 + m_{02}\sinh\alpha_2 = 0 \quad (50)$$

The relation (48) with condition (50) and the condition $\psi = \alpha_2 - \alpha_1$ can be written as

$$m_{01}\cosh\alpha_1 + m_{02}\cosh\alpha_2 = M_S \quad (51)$$

We shall write relation (51) using slopes $K\alpha_1 = \tanh\alpha_1$ and $K\alpha_2 = \tanh\alpha_2$

$$\frac{m_{01}}{\sqrt{1-K_{\alpha 1}^2}} + \frac{m_{02}}{\sqrt{1-K_{\alpha 2}^2}} = M_S \quad (52)$$

where $M_S = OO_1 = m_1 + m_2$.

$$m_1 = m_{01}\cdot\cosh\alpha_1 = \frac{m_{01}}{\sqrt{1-K_{\alpha 1}^2}} = OE, \quad m_2 = m_{02}\cdot\cosh\alpha_2 = \frac{m_{02}}{\sqrt{1-K_{\alpha 2}^2}} = EO_1 \quad (53)$$

$m_{01} = OA$, $m_{02} = O_1B$ . These transformations cause the point Q' to be transferred to Q along the circle with the center O and the radius OA = $m_{01}$. The circle QB with the center at the point $O_1$ marks the value $m_{02}$ on the same axis. The angles $\alpha_1$ and $\alpha_2$ are invariants of these transformations, in addition to $m_{01}$ , $m_{02}$ , $M_S$ . As a result, all constituents of the mass $M_S$, $m_1$, $m_2$,$m_{01}$, $m_{02}$, and $\Delta m$ = AB are invariants or functions of invariants. According to equations (52) and (53), the mass $M_S$ of the system is the sum of the relativistic masses $m_1$ *and* $m_2$ of the moving particles inside the system. A mass $\Delta m$=AB enters into its composition, in addition to the masses $m_{01}$ and $m_{02}$, as a result of the particles' motion inside the system. This is the bonded kinetic mass for the system. The components of the equation (50) are also invariants. However, the physical meaning of these components is the subject of the dynamics. The components of the equations (46) and (52) are similar with respect to the structure, but in (46) they are coordinates of the transforming vector, and in (52) they are invariants of the same Lorentz transformations.

## 9. Conclusion

The group analysis of the set of straight lines in the Minkowski plane forces us to the determination of the subgroups over sets of equivalent straight lines, corresponding to the invariants $i_1 = 0$ and $i_3 = h\boldsymbol{i}$. With $|K_x|< 1$ there are the world lines of the inertial motion. The first subgroup is a bundle over a set of straight lines with the center in the (0,0) point. The common bundle characteristic there is the unit vector $\bar{u}$, the relativistic velocity.

The analysis of the geometrical meaning of the invariant $i_3$ showed that the set of the world lines that do not contain the point (0,0) is divided into sub-groups composed of tangents to the concentric Minkowski circles with a common center at (0,0) and radius $h\boldsymbol{i}$. Within each subgroup all straight lines are equivalent with respect to the Lorentz group. The representative vector $\bar{P} = h\bar{u}$ is the geometrical characteristic of each subgroup.

Consideration of the physical meaning of the invariant $h$ forces us to the analysis of the system of two particles. The analyzed geometrical object is formed by a pair of corresponding world lines that are characterized by three invariants $h_1$, $h_2$ and the angle $\psi$ between the straight lines. This leads us to the conception of the representative vector of the system $\bar{P}_S = h_S\bar{u}_S$. The analysis of elastic interaction has led us to the conclusion that the rest mass $m_0$ of the moving particle corresponds to the invariant $h$ in physics. Thus $\bar{P} = m_0\bar{u}$ is the relativistic momentum of the particle.

The investigation of the properties of the mass of a system of particles has forced us to consider the hyperbolic triangle as a geometric object. It is formed by the vectors $\bar{P}_1$ , $\bar{P}_2$ and their sum $\bar{P}_S$. Here the object is characterized by four invariants $h_1, h_2$ and the two angles between the vectors $\bar{P}_1$ , $\bar{P}_S$ and $\bar{P}_2$ , $\bar{P}_S$. The analysis of this geometrical object leads us to the formulation of the proper system of relations, to the mass disintegration into the additive components, and to the mass–velocity dependence. Analysis shows that an increase of mass with an increase of velocity takes place only in a system of particles in which the mass of the system contains not only the masses of the particles but also the bonded kinetic mass.

## References

Gel'fand, I. M., Minlos, R. A., Shapiro, Z. Ya.: *Representations of the rotation and Lorentz groups and their applications*. Pergamon Press, Oxford **(1963)**

Naimark, M. A.: *Linear representations of the Lorentz group*. Pergamon Press, New York **(1964)**

Rashevsky, P. K.: *Riemannian geometry and tensor analysis*. Nauka, Moscow **(1967)**

McConnell, A. J.: *Application of tensor analysis*. Dover, New York **(1957)**